\documentclass[12pt,english,floatfix,showkeys,superscriptaddress,aps,prd,preprint]{revtex4}
\usepackage[latin1]{inputenc}
\usepackage[T1]{fontenc}
\usepackage{lmodern}
\usepackage{amsmath}
\usepackage{amssymb}
\usepackage{graphicx}
\usepackage{float}
\usepackage{esint}
\usepackage{longtable}
\usepackage{dcolumn}
\usepackage{babel}
\usepackage{csquotes}
\usepackage{color}
\usepackage{slashed}
\usepackage{simplewick}
\usepackage{amsmath,latexsym}

\usepackage{hyperref}
\hypersetup{
    colorlinks,
    citecolor=blue,
    filecolor=green,
    linkcolor=purple,
    urlcolor=red,
}

\usepackage{slashed}

\usepackage{hyperref}
\hypersetup{colorlinks,breaklinks,
			citecolor=[rgb]{0,0.0,1.0},
            urlcolor=[rgb]{0.0,0.0,1.0},
            linkcolor=[rgb]{0,0.5,0.9}}

\begin{document}

\title{ Generating 4-dimensional Wormholes with Yang-Mills Casimir Sources}

\author{A. C. L. Santos}
\email{alanasantos@fisica.ufc.br}
\affiliation{Universidade Federal do Cear\'a (UFC), Departamento de F\'isica,\\ Campus do Pici, Fortaleza - CE, C.P. 6030, 60455-760 - Brazil.}
\author{R. V. Maluf}
\email{r.v.maluf@fisica.ufc.br}
\affiliation{Universidade Federal do Cear\'a (UFC), Departamento de F\'isica,\\ Campus do Pici, Fortaleza - CE, C.P. 6030, 60455-760 - Brazil.}

\author{C. R. Muniz}
\email{celio.muniz@uece.br}
\affiliation{Universidade Estadual do Cear\'a (UECE), Faculdade de Educa\c{c}\~ao, Ci\^encias e Letras de Iguatu, Av. D\'ario Rabelo s/n, Iguatu-CE, 63.500-000 - Brazil.}


\date{\today}

\begin{abstract}
This work presents a new static and spherically symmetric traversable wormhole solution in General Relativity, which is supported by the quantum vacuum fluctuations associated with the Casimir effect of the Yang-Mills field confined between perfect chromometallic mirrors in $(3+1)$ dimensions, recently fitted using first-principle numerical simulations. Initially, we employ a perturbative approach for $x = m r \ll 1$, where $m$ represents the Casimir mass and $r$ is the radial coordinate. This approach has proven to be a reasonable approximation when compared with the exact case in this regime. To find well-behaved redshift functions, we impose constraints on the free parameters. As expected, this solution recovers the electromagnetic-like Casimir solution for $m = 0$. Analyzing the traversability conditions, we graphically find that all are satisfied for $ 0 \leq m \leq 0.17$. On the other hand, all the energy conditions are violated, as usual in this context due to the quantum origin of the source. Stability from Tolman-Oppenheimer-Volkov (TOV)  equation is guaranteed for all $r$ and from the speed of sound for $0.16 \leq m \leq 0.18$. Therefore, for $0.16 \leq m \leq 0.17$, we will have a stable solution that satisfies all traversability conditions.
\end{abstract}

\keywords{General Relativity. Yang-Mills Field. Casimir Effect. Wormholes.}

\maketitle


\section{Introduction}
Some inconsistencies within the framework of General Relativity have demanded severe efforts from physicists. Singularities and the incompatibility with observations in the cosmological scenario, among others, have guided different proposals: adding new invariants or fields, exploring alternative geometries and quantization (a great compilation can be found in \cite{CANTATA:2021ktz}). However, faced with these problems, we inquire: Are we adequately comprehending and describing the vacuum structure? What effects could arise when we include experimentally confirmed quantum vacuum effects?

One of the most established effects is called the Casimir effect and is associated with quantum vacuum fluctuations when we impose boundary conditions \cite{Casimir:1948dh}. Over the years, studies have emerged to investigate the effect of curved spaces on Casimir energy density \cite{Sorge:2019kuh, Santos:2020taq, Santos:2021jjs, Mota:2022qpf}. On the other hand, following Garattini's work, this energy density and the associated pressure was incorporated as source in Einstein's equations, resulting in the formation of wormholes for the Casimir effect of the electromagnetic field in $(3+1)$ \cite{Garattini:2019ivd}, $(2+1)$ \cite{Alencar:2021ejd} and $D$ dimensions \cite{Oliveira:2021ypz}. Also with the Casimir effect of the Yang-Mills field in $(2+1)$ dimensions \cite{Santos:2023zrj} and  extensions of General Relativity \cite{Cruz:2024ihb,Hassan:2022ibc,Zubair:2023abs,Hassan:2022hcb,Tripathy:2020ehi,Azmat:2023ygn,Mishra:2023bfe}.

Wormholes are hypothetical structures that would connect two distinct regions of spacetime. However, unlike Black Holes, which already have observational confirmation and well-established sources provided by stellar evolution \cite{EventHorizonTelescope:2019ths, EventHorizonTelescope:2022wkp}, wormholes suffer from a series of questions, including a natural formation process. In 1988, Morris and Thorne conducted a detailed study of the characteristics this hypothetical structure would need to possess in order to be traversable \cite{Morris:1988cz}. They concluded that it would require negative energy density - exotic matter - which at the time increased the skepticism related to this solution. Interestingly, one of the characteristics of the Casimir energy density is that for certain configurations, it can assume negative values. This is why it has been proposed as a potential source for this solution.

In this sense, based on the previously mentioned results, this work aims to restate the possibility of a wormhole formation with a new source given by the Casimir energy density and the corresponding pressure of the Yang-Mills field in $(3+1)$ dimensions. This quantity was recently obtained through first-principles numerical simulations in Lattice QCD, where it was identified that the Casimir interaction between perfect chromometallic mirrors reveals the presence of a new gluonic state with a mass of $m = 0.49(5)$ GeV \cite{Chernodub:2023dok}. 
On the other hand, due to the numerical difficulties in treating the solution exactly and taking into account that this effect occurs at microscopic distances for small masses, on considering a gravitational scenario we will make a perturbative approach for $x = mr \ll 1$, and to eliminate the singularities that arise at the throat - common in this context - we will impose a constraint on the free parameters.

Our paper is organized as follows: In section \ref{section II}, considering a perturbative approach, we find the shape and redshift functions, analyzing how the Casimir mass influences the wormhole characteristics. In section \ref{section III}, to identify the physical consistency of this solution, we investigated the traversability conditions, energy conditions, stability from the speed of sound, and TOV equation. Finally, in section \ref{conclusion} we outline our conclusions.

\section{Wormhole Solution}\label{section II}
\begin{figure}[!]
\begin{center}
\includegraphics[height=7cm]{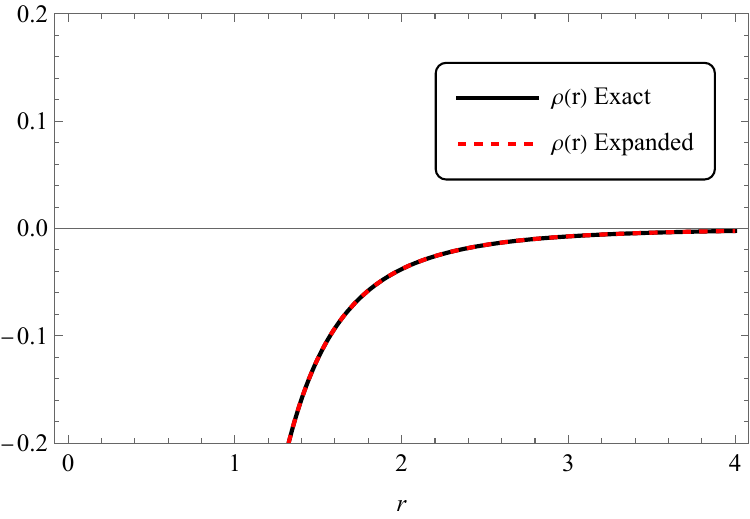}
\caption{The graphical representation of the radial dependence for $\rho(r)$ exact and expanded, with $n=1,...,100,  C_0 = 5.6$ and $ m = 5$x$10^{-2}$. }\label{Figure1}
\end{center}
\end{figure}

Since we are going to consider General Relativity, we begin by defining the well-known Einstein-Hilbert action in $(3+1)$ dimensions
\begin{equation}\label{action}
S= \frac{1}{16 \pi}\int d^4x\sqrt{-g}( R + \mathcal{L}_m),    
\end{equation}
where $g$ stands for the determinant of the metric $g_{\mu\nu}$, $R$ is the Ricci scalar and $\mathcal{L}_m$ is the Lagrangian density of matter. Varying (\ref{action}) concerning the metric, we obtain the equations of motion
\begin{equation}\label{EFE}
R^\mu_\nu -\frac{1}{2} g^\mu_\nu R = \kappa T^\mu_\nu,    
\end{equation}
with $\displaystyle \kappa = 8\pi$, where $G=\hbar = c = 1$. Assuming that the Casimir energy density and pressures effectively act as a fluid with density $\rho (r)$, radial pressure $p_r (r)$, and tangential pressure $p_t(r)$, we will consider:
\begin{equation}\label{TEM}
T_{\nu}^{\mu}=\mbox{diag (-\ensuremath{\rho (r),p_{r} (r), p_t (r),p_t (r)})}.  
\end{equation}
Let us assume a spherically symmetric and static ansatz for the spacetime metric with the line element given by
\begin{equation}\label{ansatz}
ds^{2}=-e^{2\Phi(r)}dt^{2}+\frac{1}{1-\frac{b(r)}{r}}dr^{2}+r^{2}d\theta^{2} + r^{2}\sin^2\theta d\phi^2,    
\end{equation}
which represents a $(3+1)$-dimensional Morris-Thorne wormhole \cite{Morris:1988cz}, where the redshift function $\Phi(r)$ and the shape function $b(r)$ are arbitrary functions of the polar coordinate $r\in\left[r_{0},+\infty\right)$. Thus, the coordinate $r$ must be decreased from infinity to a minimum value $r_{0}$, the radius of the throat. Replacing the ansatz to the metric (\ref{ansatz}) and the energy-momentum tensor (\ref{TEM}) in the Einstein field equations (\ref{EFE}), we find
\begin{equation}\label{eq1}
\frac{b'(r)}{r^2} = \kappa \rho(r),    
\end{equation}
\begin{equation}\label{eq2}
\kappa p_r(r) = -\frac{b(r)}{r^3} + \frac{2\Phi'(r)}{r} - \frac{2b(r)\Phi'(r)}{r^2}, 
\end{equation}
\begin{eqnarray}\label{eq3}
\left(1-\frac{b(r)}{r}\right)\left( \Phi''(r) + \Phi'^2(r) + \frac{\Phi'(r)}{r} \right) - \frac{b'(r)r - b(r)}{2r^2}\left(\Phi'(r) + \frac{1}{r}\right) = \kappa p_t(r).    
\end{eqnarray}
The prime $(')$ stands for the total derivative concerning the radial coordinate $r$. On the other hand, the covariant energy-momentum conservation law leads to
\begin{equation}\label{conservation}
p_r'(r) = \frac{-2p_r(r) + 2p_t(r) - r (p_r(r) + \rho(r))\Phi'(r)}{r}.    
\end{equation}
We will consider as a source the Casimir energy in non-Abelian gauge theory described as the Casimir energy of a massive scalar particle with certain mass $m$ which was recently fitted using first-principles numerical simulations \cite{Chernodub:2023dok} 
\begin{equation}\label{ecas}
\frac{\mathcal{E}_{Cas}}{S} = \sum _{n=1}^{\infty } - \frac{2C_0m^2}{\pi^2r} \frac{K_2(2 n m r)}{n^2},  
\end{equation}
where $S$ is the surface of the plates, $C_0$ is a phenomenological parameter, $r$ is the distance between the plates and $\displaystyle K_2(z)$ gives the modified Bessel function of the second kind. The analysis provides the following best-fit parameters: $C_0 =  5.60(7)$ and $m = 0.49(5)$ GeV.  Due to the numerical difficulties encountered when dealing with this expression exactly, we will use the perturbative approach considering $\displaystyle x = m r$, for $x\ll1$ \cite{Maluf:2019ujv}.  In this limit, the asymptotic behavior of the modified Bessel function for a fixed value of $n$ can be described by \cite{Abramowitz}:
\begin{equation}
K_2(2nmr) = \frac{2}{(mnr)^2} - \frac{1}{2} + O(m^2).   
\end{equation}
 Substituting into the summation over $n$, we obtain the following Casimir energy \cite{Plunien:1986ca}: 
\begin{equation}\label{energia}
\frac{\mathcal{E}_{Cas}(m\ll r^{-1})}{S} = -\frac{\pi^4 \lambda}{180r^3} + \frac{15\pi^2\lambda x^2}{180r^3} +O(x^4),
\end{equation}
where,
\begin{equation}
\lambda = \frac{2C_0}{\pi^2}.    
\end{equation}In which the first term is associated with the Casimir energy of a massless scalar field. Finally we can obtain the Casimir energy density
\begin{equation}\label{density}
\rho(m\ll r^{-1}) = \frac{\mathcal{E}_{Cas}(m\ll r^{-1})}{Sr} = -\frac{\pi ^2 \lambda  \left(\pi ^2-15 m^2 r^2\right)}{180 r^4} .
\end{equation}
To confirm the accuracy of this density as an approximation of the exact function within this regime, we generated a comparative graph depicted in Figure \ref{Figure1}. This graph juxtaposes the exact densities with those expanded using the best fit parameters $C_0$ and $m=5\times10^{-2}$ \cite{Chernodub:2023dok}. Through this comparison, we can confidently assert its validity as a reliable approximation.

\begin{figure}[!]
\begin{center}
\includegraphics[height=7cm]{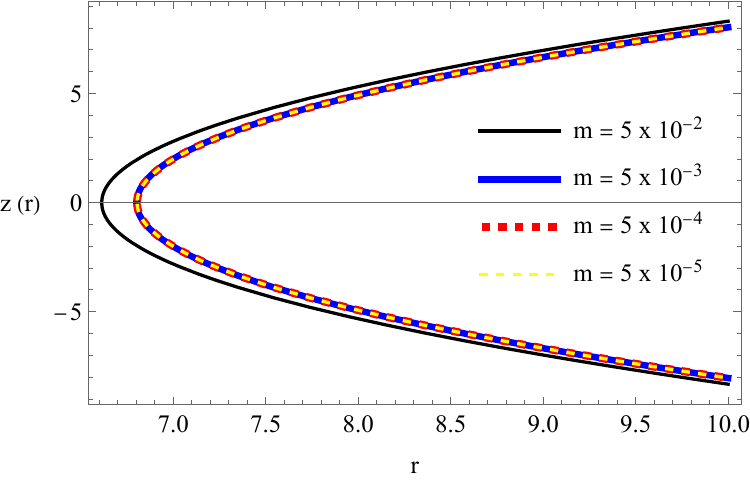}
\caption{Representation of the embedding diagram with different values of $m$, with $C_0 = 5.6$ and $r_0$ given by (\ref{constraint}), in natural units where $G = \hbar = c =1$. }\label{FigureE1}
\end{center}
\end{figure}

\begin{figure}[!h]
\begin{center}
\begin{tabular}{ccc}
\includegraphics[height=6cm]{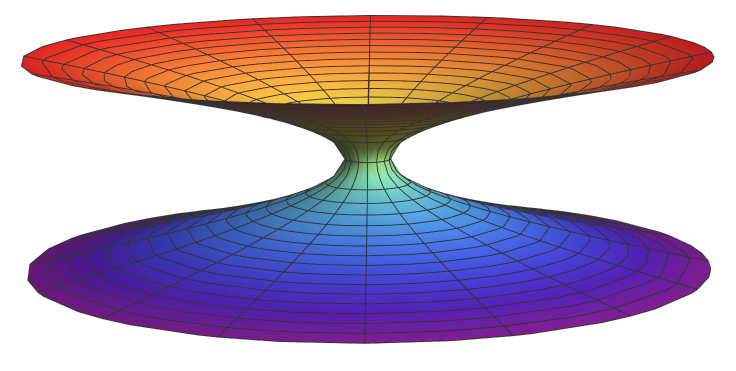}
\end{tabular}
\end{center}
\caption{Embedded shape of the (3+1) Casimir-Yang-Mills wormhole with $ m = 5$x$10^{-2}, C_0 = 5.6$ and $r_0$ given by (\ref{constraint}), in natural units where $G = \hbar = c =1$.}\label{Wormhole}
\end{figure}
From (\ref{eq1}) and (\ref{density}) we obtain the shape function
\begin{eqnarray}\label{B(r)1}
b(m\ll r^{-1}) = \frac{\pi ^2 \kappa  \lambda  (r-r_0) \left(15 r_0 m^2 r - \pi ^2\right)}{180 r_0 r}+r_0.
\end{eqnarray}
The constant of integration was fixed such that the throat condition $b(r_0)=r_0$ is satisfied. When $m \to 0$, we recover the profile of the wormhole solution sustained by the Casimir effect of the electromagnetic field, as expected \cite{Garattini:2019ivd}. Besides this, we observe that in the limit $r \to \infty $ the expression
\begin{equation}
\frac{b(r)}{r}\approx \frac{\pi^2\kappa\lambda m^2 }{12},
\end{equation}
indicates that the spacetime exhibits a defect in the solid angle, similar to the behavior of a global monopole, for sufficiently large distances from the throat. This characteristic arises due to the non-trivial topology introduced by the quantum fluctuations of the Yang-Mills field, which modifies the angular structure of the spacetime and leads to a deficit angle. Note that this feature does not occur when $m\to 0$, corresponding to the usual electromagnetic limit.

Analyzing Figure \ref{FigureE1}, which depicts embedding diagrams for different values of $m$, we identified that the larger the Casimir mass, the smaller the throat radius, which aligns with the constraint given by the equation (\ref{constraint}) and can be interpreted as an increase in $m$ leading to more intense quantum effects, causing pronounced distortions in the geometry of the wormhole. However, very tiny masses have a throat radius that is very close due to the dependence of $r_0$ with the inverse of $m^2$. Figure \ref{Wormhole} represents this three-dimensional embedding, highlighting its asymptotic flatness. 

\begin{figure}[!]
\begin{center}
\includegraphics[height=7cm]{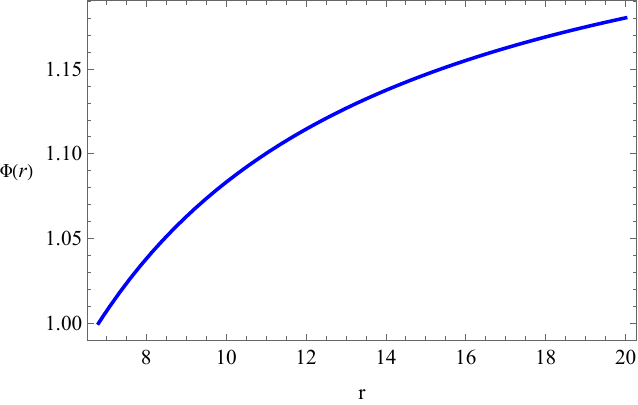}
\end{center}
\caption{ Redshift function from equation (\ref{redshift1}) with $ m = 5$ x $10^{-4}, C_0 = 5.6$ and $r_0$ given by (\ref{constraint}), in natural units where $G = \hbar = c =1$.\label{Redshift}}
\end{figure}

\begin{figure}[!]
\begin{center}
\includegraphics[height=7cm]{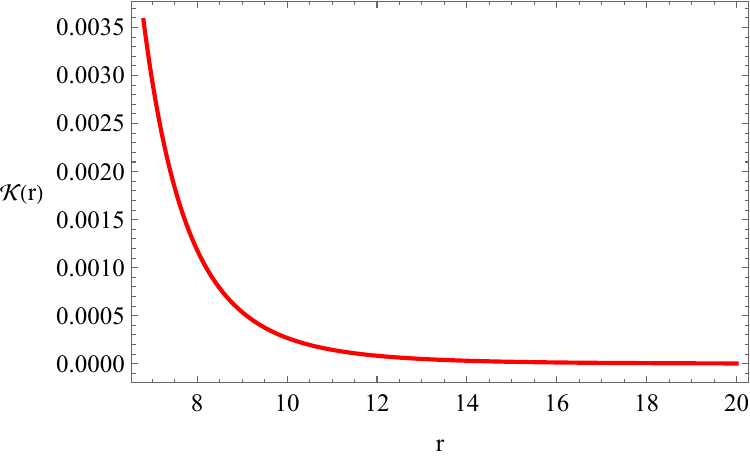}
\end{center}
\caption{Kretschmann scalar with $ m = 5$x$10^{-4}, C_0 = 5.6$ and $r_0$ given by (\ref{constraint}), in natural units where $G = \hbar = c =1$.\label{Kretchsmann}}
\end{figure}

 Let us turn our attention to the redshift function associated with the Casimir-Yang Mills wormhole. The radial pressure is given by
\begin{eqnarray}\label{pression}
p_{r}(m\ll r^{-1}) = -\frac{1}{S}\frac{d \mathcal{E}_{Cas}}{dr} = -\frac{\pi ^2 \lambda  \left(\pi ^2-5 m^2 r^2\right)}{60 r^4}.
\end{eqnarray}
This allows us to conclude the following equation of state:
\begin{equation}
p_r(m\ll r^{-1}) = \omega (m\ll r^{-1}) \rho (m\ll r^{-1});\;\;\omega (m\ll r^{-1}) = \frac{2 \pi ^2}{\pi ^2-15 m^2 r^2}+1,
\end{equation}
which recovers the $\omega$ of the electromagnetic case for $m=0$, as expected. From (\ref{eq2}), (\ref{B(r)1}) and (\ref{pression}) we obtain
\begin{eqnarray}\label{A1}
&&\Phi(m\ll r^{-1}) = \frac{1}{2 \left(\pi ^2 \kappa  \lambda  m^2-12\right) \left(\pi ^4 \kappa  \lambda -15 {r_0}^2 \left(\pi ^2 \kappa  \lambda  m^2-12\right)\right)} \nonumber \\
&&\times \left(3 \ln (r-r_0) \left(\pi ^2 \kappa  \lambda  m^2-12\right) \left(5 {r_0}^2 \left(\pi ^2 \kappa  \lambda  m^2+12\right)-\pi ^4 \kappa  \lambda \right) \right. \nonumber \\
&&+\left.\left(45 {r_0}^2 \left(\pi ^2 \kappa  \lambda  m^2-12\right)^2-\pi ^4 \kappa  \lambda  \left(\pi ^2 \kappa  \lambda  m^2+12\right)\right) \ln \left(15 r_0 r \left(\pi ^2 \kappa  \lambda  m^2-12\right)-\pi ^4 \kappa  \lambda \right)\right) \nonumber \\
&&+ \ln (r) + c_1,
\end{eqnarray}
where $c_1$ is a constant. In order to have a traversable wormhole we need to eliminate the divergence in the logarithmic term at $r \to r_0$, which implies the constraint
\begin{equation}\label{cII}
5 {r_0}^2 \left(\pi ^2 \kappa  \lambda  m^2+12\right)-\pi ^4 \kappa  \lambda = 0,    
\end{equation}
 which leads
 \begin{eqnarray}\label{constraint}
r_0 = \pi ^2 \sqrt{\frac{\kappa\lambda}{5 \pi ^2 \kappa  \lambda  m^2+60}}.
\end{eqnarray}
Considering (\ref{cII}), the redshift function becomes:
\begin{eqnarray}\label{redshift1}
\Phi(m\ll r^{-1}) &=& \frac{2 \left(\pi ^2 \kappa  \lambda  m^2-6\right) \ln\left(10 {r_0}^2 \left(\pi ^2 \kappa  \lambda  m^2-24\right)\right)}{\pi ^2 \kappa  \lambda  m^2-12}+\ln \left(\frac{r}{r_0}\right)+1 \nonumber \\
&-&\frac{2 \left(\pi ^2 \kappa  \lambda  m^2-6\right) \ln \left(-5 \pi ^2 r_0 \kappa  \lambda  m^2 (r_0-3 r)-60 r_0 (r_0+3 r)\right)}{\pi ^2 \kappa  \lambda  m^2-12},
\end{eqnarray}
which is a finite quantity when $r \to r_0$. The constant of integration $c_1$ was fixed such that $\Phi(r_0)=1$ is satisfied. This behavior is illustrated in Figure \ref{Redshift}. Note that for the argument of the first logarithm we have the condition $\pi^2\kappa\lambda m^2 \neq 24$ and for the second logarithm to be null, we need
\begin{equation}
5 \pi ^2 r_0 \kappa  \lambda  m^2 (3r- r_0)-60 r_0 (r_0+3 r)=0,    
\end{equation}
which implies,
\begin{equation}
r = r_0 \left(\frac{8}{\pi ^2 \kappa  \lambda  m^2-12}+\frac{1}{3}\right) < r_0.
\end{equation}
It is valid to mention that for $m \to 0$, we recover the solution profile sustained by the electromagnetic Casimir source \cite{Garattini:2019ivd}. Certainly, the redshift function will diverge as $r \to \infty$ due to the presence of the second logarithmic term. However, in this limit, the approximation $x = mr \ll 1$ is no longer valid. Therefore, we cannot use (\ref{density}) to describe the Casimir energy density and, consequently, (\ref{redshift1}) as a solution to the redshift function. Since the Kretschmann scalar is an involved expression, to ensure the absence of singularities in this spacetime for this regime we plotted Figure \ref{Kretchsmann}.


\section{Source Properties}\label{section III}
\subsection{Traversability Conditions}
\begin{figure}[!]
\begin{center}
\includegraphics[height=6cm]{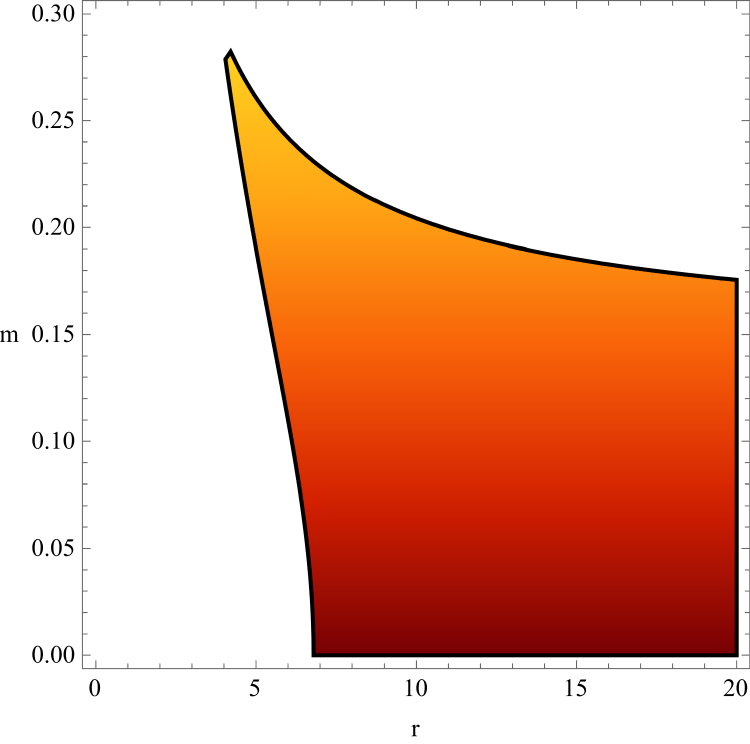}
\includegraphics[height=6cm]{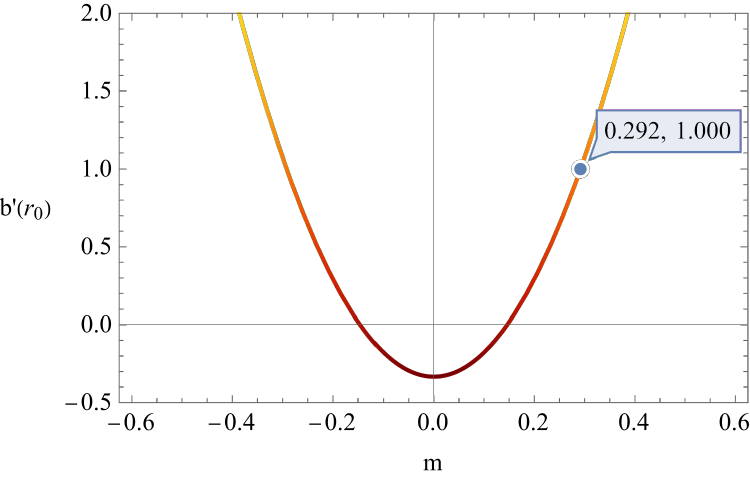}\\ 
(a) \hspace{7 cm}(b)\\
\end{center}
\caption{Region where the traversability conditions given by (\ref{c11}), (\ref{c13}) (a) and  (\ref{c12}) (b) are satisfied with $C_0 = 5.6$ and $r_0$ given by (\ref{constraint}), in natural units where $G = \hbar = c = 1$.\label{TCI}}
\end{figure}

Let's investigate whether the conditions established for traversability \cite{Morris:1988cz} are satisfied by the redshift and shape functions found. \\
$(i)$ A flaring-out condition, associated with the minimality of the wormhole throat, implies that
\begin{equation}\label{c11}
\frac{b(r) - rb'(r)}{b(r)^2} = -\frac{180 r_0 r \left(-15 {r_0}^2 r \left(12-\pi ^2 \kappa  \lambda  m^2\right)-2 \pi ^4 r_0 \kappa  \lambda +\pi ^4 \kappa  \lambda  r\right)}{\left(180 {r_0}^2 r+\pi ^2 \kappa  \lambda  (r_0-r) \left(\pi ^2-15 r_0 m^2 r\right)\right)^2} >0.    
\end{equation}
And on the throat
\begin{equation}\label{c12}
b'(r_0) = \frac{1}{180} \pi ^2 \kappa  \lambda  \left(15 m^2-\frac{\pi ^2}{{r_0}^2}\right) < 1.  
\end{equation}
$(ii)$ Another condition is given by 
\begin{equation}\label{c13}
 \displaystyle 1-\frac{b(r)}{r} = \frac{(r-r_0) \left(15 r_0 r \left(12 - \pi ^2 \kappa  \lambda  m^2\right)+\pi ^4 \kappa  \lambda \right)}{180 r_0 r^2} \geq 0.
\end{equation}
We plotted a graph in Figure \ref{TCI} to analyze which values of $m$ allow the conditions (\ref{c11}), (\ref{c12}), and (\ref{c13}) to be satisfied within the approach. Thus, we can conclude that for $ 0 \leq m \leq 0.17$, all traversability conditions will be satisfied, which is consistent with the approximation we are employing.
On the other hand, the shape function is asymptotically flat-like, since
\begin{equation}
\lim_{r\rightarrow\infty} \left(1- \frac{b(r)}{r}\right) = 1-\frac{\pi ^2 \kappa  \lambda  m^2}{12}.
\end{equation}
Despite exhibiting the desired behavior, as mentioned earlier, the asymptotic limit $r \to \infty$ does not fit within the approximation we are using. Finally, from (\ref{eq3}), (\ref{B(r)1}) and (\ref{A1}) and imposing the constraint (\ref{constraint}), we can obtain the tangential pressure:
\begin{eqnarray}\label{tanpressure}
p_t (m\ll r^{-1}) &=& -\frac{\pi ^2 \lambda  \left(3 \pi ^2 r \left(5 \kappa  \lambda  m^2 \left(\pi ^2-3 m^2 r^2\right)-36\right)- \alpha\left(15 m^2 r^2+\pi ^2\right)\right)}{180 r^4 \left(\alpha+r \left(36-3 \pi ^2 \kappa  \lambda  m^2\right)\right)},
\end{eqnarray}
where,
\begin{equation}
\alpha = \pi ^2 \sqrt{\frac{\kappa \lambda (\pi ^2 \kappa  \lambda  m^2+12)}{5}}.    
\end{equation}
It is worth mentioning that the same result is found using the equation (\ref{conservation}), as expected. Therefore, the Casimir source that generates a wormhole effectively acts as an anisotropic fluid, since the radial and tangential pressures are distinct as evidenced by the the equations (\ref{pression}) and (\ref{tanpressure}). This is in agreement with analysis in more general contexts \cite{Kim:2019ojs}.
\subsection{Energy Conditions}
The energy conditions state that for a given fluid with density $\rho(r)$, radial pressure $p_r(r)$, and lateral pressures $p_t(r)$, the following relationships must be satisfied: $\rho (r) \geq 0 $ and $\rho(r) + p_{i}(r) \geq 0$ (Weak Energy Condition), $\rho(r) + p_{i}(r) \geq 0$ (Null Energy Condition), $\rho (r) + \sum_i p_{i}(r) \geq 0 $ and $\rho(r) + p_{i}(r) \geq 0$ (Strong Energy Condition) and $\rho(r) - |p_{i}(r)| \geq 0$ (Dominant Energy Condition). 

\begin{figure}[!h]
\begin{center}
\begin{tabular}{ccc}
\includegraphics[height=5cm]{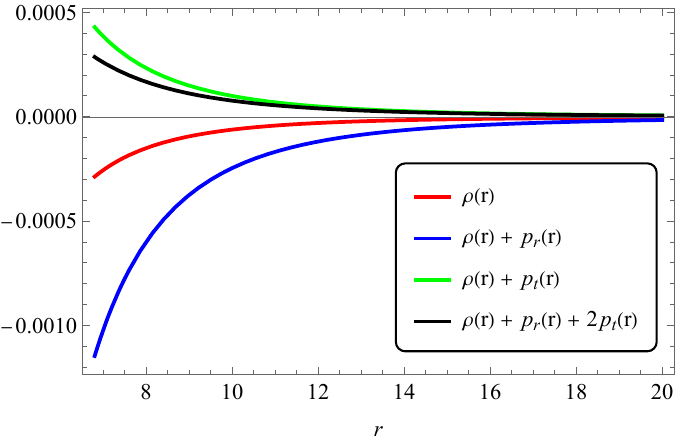}
\includegraphics[height=5cm]{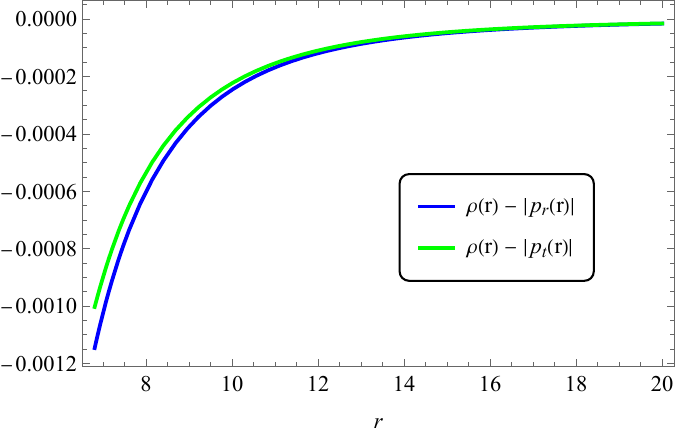}\\ 
(a) \hspace{6 cm}(b)
\end{tabular}
\end{center}
\caption{The graphical representation of the radial dependence for a combination of density and pressures (a) and (b) with $r_0$ given by (\ref{constraint}), $ m = 5$ x $10^{-4}$ and $C_0 = 5.6$, in natural units where $G = \hbar = c = 1$.\label{Figure3}}
\end{figure}

As we can see in Figure \ref{Figure3}, all the energy conditions will be violated. This fact is common in other contexts of wormholes sustained by the Casimir source and is reasonable since these conditions have a classical nature, whereas the analyzed source has a quantum origin.
\subsection{Stability from sound velocity}
\begin{figure}[!]
\begin{center}
\includegraphics[height=7cm]{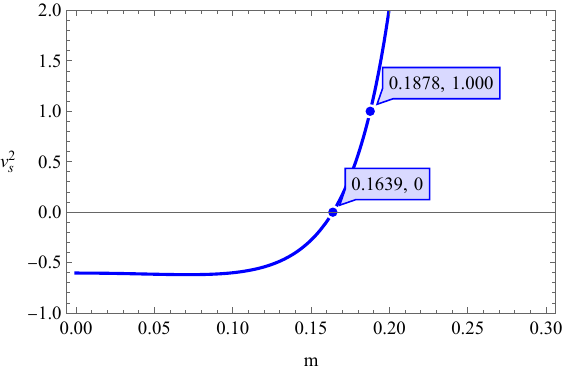}
\end{center}
\caption{Graphical representation of $v^2_s$ with $r \approx r_0$ given by (\ref{constraint}) and $C_0 = 5.6$, in natural units where $G = \hbar = c = 1$.\label{Figure5}}
\end{figure}

In order to evaluate the stability of the Casimir wormhole, we must first analyze the condition described by the expression \cite{Capozziello:2020zbx, Capozziello:2022zoz}:
\begin{equation}
v_s^2(r)=\frac{1}{3}\left[\frac{d(p_r+2p_t)}{d\rho}\right]=\frac{1}{3}\left[\frac{p_r'(r)+2p_{t}'(r)}{\rho'(r)}\right] \geq 0,\label{Sound}
\end{equation}
with $v_s$ representing the sound velocity in the medium. Considering (\ref{density}), (\ref{pression}) and (\ref{tanpressure}) we obtain
\begin{eqnarray}
v_s^2(r)=\frac{-15 m^2 r^2 + \gamma(r)}{3 \left(2 \pi ^2-15 m^2 r^2\right)} \geq 0,
\end{eqnarray}
where,
\begin{eqnarray}
\gamma (r) &=& \frac{1}{\left( \pi ^2 \sqrt{5(\kappa \lambda \pi ^2 \kappa  \lambda  m^2+12)}-15 r \left(\pi ^2 \kappa  \lambda  m^2-12\right)\right)^2} \nonumber \\
&\times& \left(10\pi ^2\left(45 \sqrt{5\kappa\lambda } m^2 r^3 \left(\pi ^2 \kappa  \lambda  m^2-18\right) \sqrt{\pi ^2 \kappa  \lambda  m^2+12}+675 \kappa  \lambda  m^4 r^4 \left(\pi ^2 \kappa  \lambda  m^2-12\right)\right. \right. \nonumber \\
&+&\pi ^4 \kappa  \lambda  \left(\pi ^2 \kappa  \lambda  m^2+12\right)+12 \pi ^2 \sqrt{5\kappa \lambda } r \left(\pi ^2 \kappa  \lambda  m^2-3\right) \sqrt{\pi ^2 \kappa  \lambda  m^2+12} \nonumber \\
&-& \left.\left.30 r^2 \left(\pi ^2 \kappa  \lambda  m^2-6\right) \left(11 \pi ^2 \kappa  \lambda  m^2-108\right)\right)\right).
\end{eqnarray}
Since stability must be analyzed in the throat, let's consider $r \approx r_0$, where $r_0$ is given by (\ref{constraint}), and examine in the Figure \ref{Figure5} which values of $m$ allow for a stable solution. As we can identify, to satisfy the condition given by (\ref{Sound}) and avoid superluminal velocities, we need $0.16\leq m \leq 0.18$. Therefore, for $0.16\leq m \leq 0.17$ we have a stable solution with the traversability conditions satisfied.

\subsection{Stability from TOV equation}
\begin{figure}[!]
\begin{center}
\includegraphics[height=7cm]{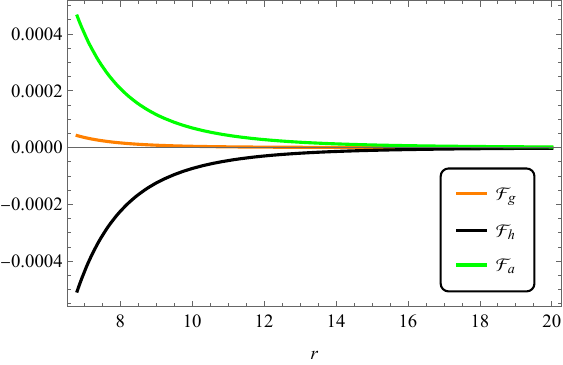}
\end{center}
\caption{The graphical representation of $\mathcal{F}_g, \mathcal{F}_h$ and  $\mathcal{F}_a$ as functions of the radial coordinate $r$, with $r_0$ given by (\ref{constraint}), $m=5$x$10^{-4}$ and $C_0 = 5.6$, in natural units where $G = \hbar = c = 1$.\label{Figure7}}
\end{figure}
From (\ref{conservation}), we can identify:
\begin{equation}
\mathcal{F}_g + \mathcal{F}_h + \mathcal{F}_a = 0,
\end{equation}
where,
\begin{equation}
\mathcal{F}_g = - \Phi'(r)(\rho(r)+p_r(r));\;\mathcal{F}_h = - \frac{dp_r(r)}{dr};\;\mathcal{F}_a = \frac{-2(p_r(r)-p_t(r))}{r},    
\end{equation}
with $\mathcal{F}_h$ the hydrostatic force, $\mathcal{F}_g$ the gravitational force and $\mathcal{F}_a$ the anisotropic force.  It is straightforward to verify that our traversable wormhole solution satisfies the TOV equation with $r_0$ given by (\ref{constraint}). The profiles of $\mathcal{F}_g$, $ \mathcal{F}_h$ and  $\mathcal{F}_a$ are depicted in Figure \ref{Figure7}. $\mathcal{F}_a$ and $\mathcal{F}_g$ take positive values near the throat, while $\mathcal{F}_h$ is negative, clearly indicating that to maintain the system in an equilibrium state, the hydrostatic force is balanced by the combined effect of gravitational
 and the anisotropic forces, in a similar behavior of $(2+1)$ dimensions \cite{Santos:2023zrj}.

\section{Conclusion \label{conclusion}}

In summary, we have found a wormhole solution supported by the energy density and Casimir pressures associated with the Yang-Mills field, which has recently been simulated using first-principles numerical simulations in $(3+1)$ dimensions. We considered a perturbative approach for $x = m r \ll 1$, where $m$ is the Casimir mass and $r$ is the radial coordinate. To eliminate the divergence of the redshift function at the throat, we imposed a constraint on its radius, $r_0$. We identified in the shape function that a larger Casimir mass implies a smaller wormhole throat. This function also demonstrates that asymptotically, the Yang-Mills Casimir wormhole exhibits a defect in the solid angle, similar to a global monopole.

In order to the traversability conditions to be satisfied, it is necessary $ 0 \leq m \leq 0.17$, which is consistent with the approximation considered. In turn, the solution is stable for all $r$ from the point of view of TOV equation, as well as of the speed sound for $ 0.16 \leq m \leq 0.18$. Thus, considering $ 0.16 \leq m \leq 0.17$ for the Casimir mass implies that we will have a stable solution that satisfies all traversability conditions. We have also shown that all the energy conditions are violated, which is typical in this context due to the quantum nature of the source. For $m=0$, we recover the expected Casimir electromagnetic solution.

To conclude, quantum vacuum fluctuations and non-linearities associated with the Yang-Mills field within hadrons could enable and stabilize submicroscopic wormholes by providing the appropriate negative energy. Although their detection remains challenging, indirect methods may uncover their presence through tiny deviations in high-energy particle collisions. Our research will seek using quasinormal modes (QNMs) to investigate the wormhole responses to perturbations, which would help guide experimental design and refine detection techniques.

\section*{Acknowledgments}
\hspace{0.5cm} The authors thank the Funda\c{c}\~{a}o Cearense de Apoio ao Desenvolvimento Cient\'{i}fico e Tecnol\'{o}gico (FUNCAP), the Coordena\c{c}\~{a}o de Aperfei\c{c}oamento de Pessoal de N\'{i}vel Superior (CAPES), and the Conselho Nacional de Desenvolvimento Cient\'{i}fico e Tecnol\'{o}gico (CNPq), Grants no. 88887.822058/2023-00 (ACLS), no. 308268/2021-6 (CRM), and no. 200879/2022-7 (RVM) for financial support.


\end{document}